# Can we implement this quantum communication ?

Tian-Hai Zeng

*Department of physics, Beijing institute of technology, Beijing 100081, People's Republic of China*

(Aug. 26, 2005)

Here I design an experimental way of a quantum communication by quantum CNOT gates and single qubit gates without the help of classical communication.

In recent years, some designs of quantum communication, for example in Ref. [1-6], should be helped by the classical communication channel, that needs sending information by electromagnetic wave or broadcast and need no sending quantum systems or qubits. It is believed that no-cloning theorem [7] prevents to distinguish non-orthogonal quantum states with perfect reliability, and prevents the quantum communication without the help of classical communication.

Here I design an experimental way of a quantum communication by quantum CNOT gates and single qubit gates without the help of classical communication. For our purpose, it is enough for a quantum gate to clone anyone state of two orthogonal states $|0>$ and $|1>$, that is, if the being copied state is $|0>$, the two outputs get two states $|0>$, and if the being copied state is $|1>$, the two outputs get two states $|1>$ by the same gate. No-cloning theorem does not prohibit this type cloning. A quantum circuit [3] has been designed for doing this type cloning. Quantum Toffoli gate is also suit for this thing. The simplest gate for this application is quantum CNOT gate. The target qubit inputs state $|0>$ and the control qubit inputs state $|0>$ or $|1>$, then two outputs get two states $|0>$ or two states $|1>$.

Now I show an experimental way for this quantum communication. Alice and Bob are far from each other, they share an EPR [8] pair $|\psi\rangle = (|00\rangle+|11\rangle)/\sqrt{2}$. Alice wants to send a bit classical information to Bob, she can select one measurement in one of the two computational bases [3]: $\{|0\rangle, |1\rangle\}$, or $\{(|0\rangle+|1\rangle)/\sqrt{2}, (|0\rangle-|1\rangle)/\sqrt{2}\}$, then Bob uses quantum CNOT gate to copy his qubit and measures his qubits only in the projection on the state $|0\rangle$. If Alice uses the first base, then Bob's qubit will be in $|0\rangle$ or $|1\rangle$ instantaneously. Bob can definitely clone his qubit using a quantum CNOT gate, No matter what state ($|0\rangle$ or $|1\rangle$) is Bob's qubit. If Alice uses the other base, then Bob's qubit will be in $(|0\rangle+|1\rangle)/\sqrt{2}$ or $(|0\rangle-|1\rangle)/\sqrt{2}$ instantaneously. Bob still uses the gate to "copy" his qubit. But this time, he *cannot get same copies,* but he *can get entangled state* $(|00\rangle+|11\rangle)/\sqrt{2}$ or $(|00\rangle-|11\rangle)/\sqrt{2}$.

To distinguish Bob's two set states $\{|0\rangle, |1\rangle\}$ and $\{(|0\rangle+|1\rangle)/\sqrt{2}, (|0\rangle-|1\rangle)/\sqrt{2}\}$, using only one quantum CNOT gate is not enough. Bob can use n-1 quantum CNOT gates or a quantum CNOT gate with n-1 target qubits to get one result of four cases: $\{|0\rangle^{(n)}, |1\rangle^{(n)}; (|0\rangle^{(n)}+|1\rangle^{(n)})/\sqrt{2}, (|0\rangle^{(n)}-|1\rangle^{(n)})/\sqrt{2}\}$, the first two states are n qubit product states and the last two states are n qubit GHZ [9] states.

Next step is pivotal. Bob knows that he has to do same things to distinguish product states and GHZ states according to his measurement results. Bob uses a single qubit gate $U^{(1)}(\theta)$ to rotate his first qubit in $\theta$ ($0<\theta<\pi/8$) [3,10] and gets:

$$U^{(1)}(\theta) |0\rangle^{(n)} = (\cos\theta|0\rangle+\sin\theta|1\rangle)|0\rangle^{(n-1)} , \qquad (1)$$

$$U^{(1)}(\theta) |1\rangle^{(n)} = (-\sin\theta|0\rangle+\cos\theta|1\rangle)|1\rangle^{(n-1)} , \qquad (2)$$

$$U^{(1)}(\theta) (|0\rangle^{(n)} + |1\rangle^{(n)})/\sqrt{2} = [|0\rangle(\cos\theta|0\rangle^{(n-1)} - \sin\theta |1\rangle^{(n-1)}) + |1\rangle(\sin\theta|0\rangle^{(n-1)} + \cos\theta |1\rangle^{(n-1)}) ]/\sqrt{2}, \qquad (3)$$

$$U^{(1)}(\theta) (|0\rangle^{(n)} - |1\rangle^{(n)})/\sqrt{2} \} = [|0\rangle(\cos\theta|0\rangle^{(n-1)} + \sin\theta |1\rangle^{(n-1)}) + |1\rangle(\sin\theta|0\rangle^{(n-1)} - \cos\theta |1\rangle^{(n-1)}) ]/\sqrt{2}. \qquad (4)$$

From Eq. (3), (4), we can get a conclusion that GHZ state is not fully fragile, as we can maintain n-1 qubit entanglement of n qubit GHZ state after having measured one of n qubits. Then Bob measures the projection of his first qubit only on state $|0\rangle$ and he can get four different probabilities:

$$p_{11}=\cos^2\theta, \quad p_{21}=\sin^2\theta, \quad p_{31}=p_{41}=1/2. \tag{5}$$

Again Bob does same things on his second qubit and easily gets probabilities of state $|0\rangle$ in the case of product states:

$$p_{12}=\cos^2\theta, \quad p_{22}=\sin^2\theta . \tag{6}$$

When Bob's GHZ state is initially in $(|0\rangle^{(n)} + |1\rangle^{(n)})/\sqrt{2}$, he can get:

$$U^{(2)}(\theta) (\cos\theta|0\rangle^{(n-1)} - \sin\theta |1\rangle^{(n-1)} ) = (\cos^4\theta+\sin^4\theta)^{1/2}|0\rangle [(\cos^4\theta+\sin^4\theta)^{-1/2} (\cos^2\theta|0\rangle^{(n-2)}$$
$$+\sin^2\theta |1\rangle^{(n-2)} )] + \sqrt{2} \sin\theta\cos\theta |1\rangle (|0\rangle^{(n-2)} - |1\rangle^{(n-2)})/\sqrt{2}, \tag{7}$$

if the measurement result of Bob's first qubit is in $|0\rangle$; and

$$U^{(2)}(\theta) (\sin\theta|0\rangle^{(n-1)} + \cos\theta|1\rangle^{(n-1)} ) = \sqrt{2} \sin\theta\cos\theta |0\rangle (|0\rangle^{(n-2)} - |1\rangle^{(n-2)} )/\sqrt{2} +$$
$$(\cos^4\theta+\sin^4\theta)^{1/2}|1\rangle[(\cos^4\theta+\sin^4\theta)^{-1/2} (\sin^2\theta|0\rangle^{(n-2)} +\cos^2\theta |1\rangle^{(n-2)} )], \tag{8}$$

if the measurement result of Bob's first qubit is not in $|0\rangle$. In this case, Bob can get the probabilities of $|0\rangle$:

$$p_{32}= \cos^4\theta+\sin^4\theta , \quad p'_{32}= 2\sin^2\theta\cos^2\theta . \tag{9}$$

When Bob's GHZ state is initially in $(|0\rangle^{(n)} - |1\rangle^{(n)})/\sqrt{2}$, he can get the probabilities same as (9)

$$p_{42}= p_{32} , \quad p'_{42}= p'_{32}.$$

After Bob measures his second qubit, he gets one of four different numbers, $2(\cos^2\theta)$, $2(\sin^2\theta)$, $1+(\cos^4\theta+\sin^4\theta)$, $0+2\sin^2\theta\cos^2\theta$, of his first two qubits in $|0\rangle$. Although Bob is difficult to distinguish the product states and GHZ states only using his first two qubits, he can use his n qubits to do same things. The calculations for the numbers of n qubit GHZ states will be very tedious. We can expect that these numbers of n qubit GHZ states different from $n(\cos^2\theta)$ and $n(\sin^2\theta)$ in calculations by properly selecting $\theta$ and n, that is if the number from Bob's measurement result tends to n(cos$^2$θ) or n(sin$^2$θ), he can identify that his n qubits are in product state, otherwise in GHZ state. Finally, Bob can get a bit classical information from Alice. In this way, Alice and Bob can do this quantum communication bypassing the barrier of no-cloning theorem and not using classical communication.

Five-photon GHZ state has been implemented in the experiment [11]. In this letter, if Bob uses 4 quantum CNOT gates, he can prepare 5 qubit product states or 5 qubit GHZ states and select proper θ to satisfy sin$^2$θ = 1/5, cos$^2$θ = 4/5. Therefore, we can think that if Bob gets one or four of his qubits in |0>, he can judge his n qubits in product state, that is corresponding to the measurement made by Alice in the base {|0>, |1>}; otherwise he can judge other result.

Some technical problems may be easily solved. For example, if it is not credible that Alice and Bob use one EPR pair to send a bit classical information, they can use several EPR pairs to send a bit classical information. Other one is that Bob cannot distinguish the two cases: no measurement and the measurement in base (|0>+|1>)/$\sqrt{2}$ and (|0>-|1>)/$\sqrt{2}$ by Alice. One method is to let Alice and Bob to measure definite EPR pairs according to definite time order. Other method is that, if the first and last EPR pairs in a group are measured in base {|0>, |1>}, then there is no need of measurement in base {(|0>+|1>)/$\sqrt{2}$, (|0>-|1>)/$\sqrt{2}$} by Alice, since the two cases are same for Bob in the sense that he can get same results of the probabilities.

In summary, I design an experimental way of this quantum communication by quantum CNOT gates and single qubit gates and show how we bypass the barrier of no-cloning theorem. This way is a pure quantum communication without the help of classical communication. It is expected that this way will be proved by experiments and further calculations.

[1] A. K. Ekert, Phys. Rev. Lett. **67**, 661 (1991).

[2] C. H. Bennett, G. Brassard, C. Crépeau, R. Jozsa, A. Peres and W. K. Wootters, Phys. Rev. Lett. **70**,

1895 (1993).

[3] M. A. Nielsen and I. L. Chuang, Quantum computation and quantum communication, (Cambridge Univ. Press, Cambridge, 2000).

[4] D. Bouwmeester, J. W. Pan, K. Mattle, M. Eibl, H. Weifurter and A. Zeilinger, Nature **390,** 575(1997).

[5] Y. H. Kim, S. P. Kulik and Y. H. Shih, Phys. Rev. Lett. **87**, 1370 (2001).

[6] C. Z. Peng, T. Yang, X. H. Bao, J. Zhang, X. M. Jin, F. Y. Feng, B. Yang, J. Yang, J. Yin, Q. Zhang, N. Li, B. L.Tian & J. W. Pan. Phys. Rev. Lett*.* **94,** 150501-1 (2005).

[7] W. K. Wootters and W. H. Zurek, Nature **299,** 802 (1982).

[8] A. Einstein, B. Podolsky and N. Rosen, Phys. Rev. **47**, 777 (1935).

[9] D. M. Greenberger, M. A. Horne and A. Zeilinger, Bell's theorem, Quantum theory, and Conceptions of the Universe, edited by M. Kafatos, Kluwer, Dordrecht, 1989.

[10] L. K. Grover, Phys. Rev. Lett. 79(2), 325 (1997).

[11] Z. Zhao, Y. A. Chen, A. N. Zhang, T. Yang, H. Briegel and J. W. Pan, quant-ph /**0402096**, Nature **430**, 54 (2004).